\documentstyle[eqsecnum,preprint,prd,aps,psfig]{revtex}
\tighten
\def\jnfont{\rm}
\def\NPB#1,{{\jnfont Nucl.\ Phys.\ }{\bf B#1},}
\def\PLB#1,{{\jnfont Phys.\ Lett.\ B }{\bf #1},}
\def\PRD#1,{{\jnfont Phys.\ Rev.\ D }{\bf #1},}
\def\PRL#1,{{\jnfont Phys.\ Rev.\ Lett.\ }{\bf #1},}
\def\ZPC#1,{{\jnfont Z.~Phys.\ C }{\bf #1},}

\begin{document}

\preprint{\parbox{2.0in}{\noindent TU-545\\ RCNS-98-11\\ AMES-HET-98-03\\
 June 1998\\}}

\title{\ \\[15mm] 
Probing anomalous top quark interactions\\[2mm]
       at the Fermilab Tevatron collider\\}

\author{Ken-ichi Hikasa$^1$, K. Whisnant$^2$, Jin Min Yang$^1$ 
and Bing-Lin Young$^2$\\\ }

\address{
$^1$ {\it Department of Physics, Tohoku University,\\
      Aoba-ku, Sendai 980-8578, Japan}\\[2mm]
$^2$ {\it Department of Physics and Astronomy, Iowa State University, \\
     Ames, Iowa 50011, USA}}

\maketitle

\begin{abstract}
We study the effects of dimension-six operators contributing to the 
$gt\bar t$ vertex in top quark pair production at the Tevatron collider.  
We derive both the limits from Run~1 data and the potential 
bounds from future runs (Run~2 and 3).  Although the current 
constraints are not very strong, the future runs are quite effective 
in probing these operators.  We investigate the possibility of disentangling 
different operators with the $t\bar t$ invariant mass distribution and 
the top quark polarization asymmetry. We also study the effects of a
different set of operators contributing to single top production via the
$Wt\bar b$ coupling.  We derive the current and potential future 
bounds on these anomalous operators and find that the upgraded Tevatron
can improve the existing constraints from $R_b$ for one of the operators.  
\end{abstract}


\section{Introduction}
\label{sec:one}

The phenomenological success of the standard model (SM) has
significantly limited the possibility of new physics.  However, some
unanswered fundamental questions suggest that the SM will be
augmented by new physics at higher energy scales.  As the most massive
fermion in the SM, the top quark is naturally regarded to be more
sensitive to new physics than lighter fermions.  
Therefore, precision measurements of top quark properties offer one of the 
best possibilities to obtain information on new physics.

The measurements of top quark properties in Run~1 at the Fermilab
Tevatron have only small statistics.  There is plenty of room for new
physics to be discovered in similar measurements taken with higher
luminosities. Such possibilities exist in the near future: Run~2 and
Run~3 of the Tevatron collider should significantly improve the
precision of the measurements of top quark properties~\cite{ref1}, and
provide a better insight to new physics at higher energy scales.

There are numerous speculations on the possible forms of new physics.
The fact that no single clear signal for deviation from the SM has been
observed in any experiment strongly suggests that the
new physics effects not too far above the electroweak scale should
preserve the SM structure and at most modify it delicately. This
suggests that any new particles which may exist will be too heavy to be
produced at current and quite possibly at near-future colliders, and
thus the only observable effects of new physics at energies not too far
above the SM energy scale could be in the form of anomalous interactions
which will slightly affect the couplings of the SM particles.  This
reasoning leads to the effective Lagrangian approach in describing new
physics effects.  For the description of new physics in the top quark
sector, there are two effective Lagrangian approaches, which are
formulated in terms of non-linear~\cite{ref2} and linear
realizations~\cite{ref3,Gounaris1,Gounaris2,ref5,ref6} of the
electroweak symmetry, corresponding to the situations without and with a
light Higgs boson, respectively. In this article, we choose the linear
realization and parameterize the new physics effects in the top quark
sector by dimension-six operators which are invariant under ${\rm
SU}(3)_c\times {\rm SU}(2)_L\times {\rm U}(1)_Y$.  We will focus on the
CP-conserving operators contributing to the anomalous $gt\bar t$ or
$Wt\bar b$ coupling which can be directly probed at the upgraded
Tevatron collider through top pair and single top production,
respectively.

The operators we consider describe the anomalous couplings of top quark
with gauge and/or Higgs bosons and do not include top quark four-fermion
($q\bar qt\bar t$) contact terms, which have been investigated
at Tevatron collider by Hill and Parke~\cite{ref7}.  
There have also been analyses of the
phenomenology of an anomalous chromo-magnetic dipole moment $gt\bar t$
coupling at the Tevatron collider~\cite{ref8}. The operators considered
in this paper will also induce such an anomalous chromo-magnetic dipole
moment coupling for $gt\bar t$, but at the same time give rise to other
kinds of anomalous couplings via operators that obey the SM symmetries.
Some of the operators contributing to the $Wt\bar b$ coupling also
induce an anomalous $Zb\bar b$ coupling.  Their effects on single top
production cross section at the Tevatron were evaluated in
Refs.~\cite{ref5} and \cite{ref9}, with constraints derived from the
earlier, less accurate data on
$R_b=\Gamma(Z\to b\bar b)/ \Gamma(Z\to\rm hadrons)$.  Our
systematical analyses in this paper also include these operators, but
with a more complete calculation by considering all possible backgrounds
and by deriving the potential limits from the upgraded Tevatron.  We
find that the operators which contribute to the $Zb\bar b$ coupling are
strongly constrained and can
no longer give rise to observable effects at the upgraded Tevatron due
to the good agreement of $R_b$ with the SM value.
 
This paper is organized as follows. In Sec.~\ref{sec:2}, we list the relevant
dimension-six operators.  In Sec.~\ref{sec:3}\ we study the contribution 
of those operators to the $gt\bar t$ coupling and their effect on top pair
production.  We derive the limits on the coupling strengths set by Run~1
and determine the limits which could be set by Run~2 and Run~3 if no new
physics effects are observed.  In Sec.~\ref{sec:4}, we present two methods 
to disentangle the effects between different operators contributing 
to the $gt\bar t$ coupling. One is studying the $t\bar t$ invariant mass 
distribution of the cross section, the other is measuring the asymmetry 
between left- and right-handed top events in top pair production. 
In Sec.~\ref{sec:5}\ we evaluate the effects of those
operators contributing to the $Wt\bar b$ coupling in single top production
and derive the bounds which could be set by Run~2 and Run~3.  Finally, 
Sec.~\ref{sec:last} is devoted to conclusions.

\section{List of the relevant operators}
\label{sec:2}

We assume here that the new physics in the quark sector only resides 
in the interaction of the third family to gauge bosons and/or Higgs boson. 
The effective Lagrangian including the new physics effects can be written 
as~\cite{ref3}
\begin{equation}
{\cal L}_{eff}={\cal L}_0+\frac{1}{\Lambda^2}\sum_i C_i O_i
                         +O\biggl(\frac{1}{\Lambda^4}\biggr)
\end{equation}
where ${\cal L}_0$ is the SM Lagrangian, $\Lambda$ is the new physics
scale, $O_i$ are dimension-six operators which are
${\rm SU}(3)_c\times {\rm SU}(2)_L\times {\rm U}(1)_Y$ invariant, and 
$C_i$ are coupling constants which represent the strengths of $O_i$.  
Recently, the effective operators involving the third-family quarks were 
reclassified in Refs.~\cite{Gounaris1,Gounaris2,ref5} for
the  CP-conserving ones and in Ref.~\cite{ref6} for CP-violating ones.
There are 10 dimension-six CP-conserving operators
which contribute to the $gt\bar t$ or  $Wt\bar b$ coupling
\begin{mathletters}
\begin{eqnarray}
O_{tG}&=&\left[ \bar t_R\gamma^{\mu}T^A D^{\nu}t_R
         +\overline{D^{\nu}t_R} \gamma^{\mu}T^A t_R\right ]
          G^A_{\mu\nu},\\
O_{qG}&=&\left [\bar q_L \gamma^{\mu}T^A D^{\nu}q_L
         +\overline{D^{\nu}q_L} \gamma^{\mu}T^A q_L\right ]
          G^A_{\mu\nu},\\
O_{tG\Phi}&=&\left [(\bar q_L \sigma^{\mu\nu}T^A t_R) \widetilde\Phi
         +\widetilde\Phi^{\dagger}(\bar t_R \sigma^{\mu\nu}T^A q_L)\right ]
          G^A_{\mu\nu},\\
\label{OqW}
O_{qW}&=&\left [\bar q_L \gamma^{\mu}\tau^I D^{\nu}q_L
         +\overline{D^{\nu}q_L} \gamma^{\mu}\tau^I q_L\right ]
          W^I_{\mu\nu},\\
O_{\Phi q}^{(3)}&=&i\left [\Phi^{\dagger}\tau^I D_{\mu}\Phi
        -(D_{\mu}\Phi)^{\dagger}\tau^I\Phi\right ]\bar q_L \gamma^{\mu}\tau^I 
       q_L,\\
O_{Db}&=&(\bar q_L D_{\mu} b_R) D^{\mu}\Phi
         +(D^{\mu}\Phi)^{\dagger}(\overline{D_{\mu}b_R}q_L),\\
O_{bW\Phi}&=&\left [(\bar q_L \sigma^{\mu\nu}\tau^I b_R) \Phi
         +\Phi^{\dagger}(\bar b_R \sigma^{\mu\nu}\tau^I q_L)\right ]
          W^I, \\
O_{t3}&=&i\left [(\widetilde\Phi^{\dagger}D_{\mu}\Phi)(\bar t_R \gamma^{\mu}b_R
         -(D_{\mu}\Phi)^{\dagger}\widetilde\Phi(\bar b_R \gamma^{\mu}t_R)
       \right ],\\
O_{Dt}&=&(\bar q_L D_{\mu} t_R) D^{\mu}\widetilde\Phi
         +(D^{\mu}\widetilde\Phi)^{\dagger}(\overline{D_{\mu}t_R}q_L),\\
\label{OtWPhi}
O_{tW\Phi}&=&\left [(\bar q_L \sigma^{\mu\nu}\tau^I t_R) \widetilde\Phi
         +\widetilde\Phi^{\dagger}(\bar t_R \sigma^{\mu\nu}\tau^I q_L)\right ]
          W^I_{\mu\nu},
\end{eqnarray}
\end{mathletters}
where  $q_L=(t_L, b_L)$ denotes the third family 
left-handed quark doublet, $\Phi$ and $\tilde{\Phi}$ are the Higgs field 
and its charge conjugate $\tilde\Phi = i\tau_2\Phi^*$, $W_{\mu\nu}$ and
$B_{\mu\nu}$ are the SU(2) and U(1) gauge boson field tensors, respectively,
in the appropriate matrix forms, $D_\mu$ denotes the appropriate 
covariant derivatives, and $T^A=\lambda^A/2$ with $\lambda^A$ 
($A=1$, $\ldots$, 8) denoting the Gell-Mann matrices.   

After electroweak symmetry breaking, the first three operators 
$ O_{tG}$, $O_{qG}$ and $O_{tG\Phi}$ 
will induce anomalous $gt\bar t$ couplings which are given by
\begin{eqnarray}
{\cal L}_{g t\bar t}&=&\frac{C_{tG}}{\Lambda^2}
  \left [\bar t\gamma^{\mu}P_RT^A \partial^{\nu}t
         +\partial^{\nu} \bar t\gamma^{\mu}P_R T^A t \right ]G^A_{\mu\nu}
\nonumber\\
& & +\frac{C_{qG}}{\Lambda^2}
\left [\bar t\gamma^{\mu}P_L T^A \partial^{\nu}t
         +\partial^{\nu}\bar t \gamma^{\mu}P_L T^A t\right ] G^A_{\mu\nu}
 +\frac{C_{tG\Phi}\,v}{\sqrt2\Lambda^2}
  (\bar t\sigma^{\mu\nu}T^A t)G^A_{\mu\nu},
\end{eqnarray}
where $v=\bigl(\sqrt2 G_{\!F})^{-1/2}$ is the SM Higgs vacuum expectation 
value.  Other operators will give rise to anomalous $Wt\bar b$ couplings
which are given by
\begin{eqnarray}
\label{wtb}
{\cal L}_{Wt\bar b}&=&\frac{C_{qW}}{\sqrt2\Lambda^2} 
        W^+_{\mu\nu}(\bar t\gamma^{\mu}P_L\partial^{\nu}b
                     +\partial^{\nu}\bar t\gamma^{\mu}P_L b)
+\frac{C_{\Phi q}^{(3)}\,g_2v^2}{\sqrt2\Lambda^2}
      W^+_{\mu}(\bar t\gamma^{\mu}P_L b)\nonumber\\
& &-\frac{C_{Db}\,g_2 v}{2\Lambda^2}
      W_{\mu}^+ (i\bar t P_R \partial^{\mu}b)
   +\frac{C_{bW\Phi}\,v}{2\Lambda^2}
      W^+_{\mu\nu}(\bar t \sigma^{\mu\nu}P_R b)
+\frac{C_{t3}\,g_2 v^2}{2\sqrt2\Lambda^2}
               W_{\mu}^+ (\bar t\gamma^{\mu}P_R b) \nonumber\\
& & +\frac{C_{Dt}\,g_2 v^2}{2\Lambda^2}
     W_{\mu}^+ (i\partial^{\mu} \bar t)P_L b
 +\frac{C_{tW\Phi}\,v}{2\Lambda^2}
        W^+_{\mu\nu}(\bar t\sigma^{\mu\nu}P_L b),
\end{eqnarray}
where $g_2$ is the weak SU(2) gauge coupling.  
Note that the operators $O_{qW}$, $O_{\Phi q}^{(3)}$, $O_{Db}$ and
$O_{bW\Phi}$ also induce $Zb\bar b$ couplings
\begin{eqnarray}\label{zbb}
{\cal L}_{Zb\bar b} &=&
\frac{C_{qW}c_W}{2\Lambda^2}
   Z_{\mu\nu}(\bar b \gamma^{\mu}P_L \partial^{\nu}b
                      +\partial^{\nu}\bar b \gamma^{\mu}P_L b)
 +\frac{C_{\Phi q}^{(3)}\,v m_Z}{\Lambda^2}  
    Z_{\mu}(\bar b\gamma^{\mu}P_L b)
\nonumber\\
& &+\frac{C_{Db}\,m_Z}{2\sqrt2\Lambda^2}
    Z^{\mu}\left [i(\partial_{\mu}\bar b b-\bar b\partial_{\mu}b)
                    -i\partial_{\mu}(\bar b \gamma_5b)\right ]
+\frac{C_{bW\Phi}\,c_W v}{2\sqrt2\Lambda^2}
   Z_{\mu\nu}(\bar b \sigma^{\mu\nu} b), 
\end{eqnarray}
where $c_W \equiv \cos\theta_W$, and thus will be subject to a
constraint from $R_b$ measurements.

\section{Top pair total cross section}
\label{sec:3}

The number of top pair events produced at the Tevatron will be different
from the SM prediction if new physics exists in the top sector.  Since
the standard model cross section for top pair production is known, here
we only need to evaluate the new physics contribution to the total cross
section.%
\footnote{We assume that the new physics only affects the SM top
quark couplings and that there is no exotic decay mode for top
quark. Therefore the branching fractions of various final states in
$t\bar t$ production are about the same as in the SM.} We can predict
the number of reconstructed top pair events in the various decay
channels (dilepton, single lepton plus jets, {\it etc.}) at the upgraded
Tevatron~\cite{ref1} by extrapolating from the Run~1 results.
We will calculate the new
physics contribution to the cross section and derive bounds on the
coupling strengths of the anomalous operators.  For illustration, we will
only present the bounds at the $2\sigma$ level.

\subsection{Anomalous contribution to top pair production cross section}

The dominant mechanisms of top quark pair production at hadron colliders
are the QCD processes of quark-antiquark annihilation and gluon-gluon
fusion.  At the Tevatron collider, the quark-antiquark annihilation
process is dominant for $m_t=175$ GeV.  In the SM, several
groups~\cite{ref10,ref11,ref12} have calculated the cross section to the
next-to-leading order by summing over soft gluons up to leading
logarithms.  More recently, soft-gluon resummation at the
next-to-leading logarithmic level has also been
performed~\cite{Bonciani}.

We calculate the new physics contribution to the quark-antiquark
annihilation process and neglect its effect in the gluon-gluon
fusion process since its contribution to the cross section is small
at the Tevatron. For the on-shell $t$ and $\bar t$, we obtain the
effective $gt\bar t$ coupling arising from the dimension-six operators
\begin{equation}\label{gtt}
\Gamma^{\mu A}_{gt\bar t}= T^A \left [\gamma^{\mu} F_V
+\Bigl(\gamma^{\mu}\gamma_5-{2m_t\over k^2}k^\mu\gamma_5\Bigr) F_A
+\frac{1}{2m_t}(p_{\bar t}-p_t)^{\mu}F_M \right],
\end{equation}
where $p_t$ and $p_{\bar t}$ are the momenta of the outgoing top quark and 
anti-quark, respectively, and 
$k=p_{\bar t}+p_t$ is the momentum of the vector boson. The form factors
are derived from the contributions of the new physics operators
\begin{mathletters}
\begin{eqnarray}
F_V(k^2) &=& 
-\frac{k^2}{2}\left[\frac{C_{tG}}{\Lambda^2}+\frac{C_{qG}}{\Lambda^2}
                    \right] +2\sqrt 2 vm_t \frac{C_{tG\Phi}}{\Lambda^2},\\
F_A(k^2) &=&
-\frac{k^2}{2}\left [\frac{C_{tG}}{\Lambda^2}-\frac{C_{qG}}{\Lambda^2}
                    \right ],\\
F_M(k^2) &=& 2\sqrt 2 vm_t \frac{C_{tG\Phi}}{\Lambda^2}.
\end{eqnarray}
\end{mathletters}
The pseudoscalar coupling proportional to $k_{\mu}\gamma_5$ is related to 
the axial vector coupling by current conservation.  It gives negligible 
contribution (proportional to the initial quark mass) to the process.  

The new physics contribution to the parton-level cross section 
for $q\bar q\to t\bar t$ to order $1/\Lambda^2$ is found to be 
\begin{equation}\label{parton}
\Delta\hat \sigma^{\rm new} (\hat s) = {2\alpha_s g_s \hat\beta\over 27\hat s}
\,\bigl [ F_V(\hat s) (3-\hat\beta^2) +F_M(\hat s) \hat\beta^2 \bigr],
\end{equation}
where $\hat s$ is the center-of-mass energy squared for the parton-level
process and $\hat\beta=(1-4m_t^2/\hat s)^{1/2}$.  
Notice that the axial vector coupling does not contribute to the total 
cross section.  We will discuss in Sec.~\ref{sec:4} how to measure this 
coupling.  

The new physics contribution to the total hadronic cross section for top 
quark pair production is obtained by 
\begin{equation}\label{eq22}
\Delta\sigma^{\rm new}_{t\bar t}= \sum_{q}\int_{\tau_0}^1 \!d\tau\, 
\frac{dL_{q\bar q}}{d\tau}\, \Delta\hat\sigma^{\rm new}(\hat s=s\tau) ,
\end{equation}
where $s$ is the $p\bar p$ center-of-mass energy squared, $\hat s$ the 
center-of-mass energy squared for the parton subprocess, and 
$\tau_0=4m_t^2/s$.  
The quantity $dL_{q\bar q}/d\tau$ is the parton luminosity defined by
\begin{equation}
\frac{dL_{q\bar q}}{d\tau}=\int^1_{\tau} \frac{dx_1}{x_1}[f^p_q(x_1,\mu)
f^{\bar p}_{\bar q}(\tau/x_1,\mu)+(q\leftrightarrow \bar q)],
\end{equation}
where $f^p_q$ denotes the quark distribution function in a proton.  

In our numerical calculation, we use the CTEQ3L parton 
distribution functions~\cite{ref13} with $\mu=\sqrt {\hat s}$.
For  $m_t=175$ GeV, we obtain the new physics contribution to order
$1/\lambda$ to the total hadronic cross section in unit of pb 
\begin{equation}\label{sigman}
\Delta\sigma^{\rm new}_{t\bar t}=\cases{ 
 \bigl[  -0.61 (C_{tG}+ C_{qG}) + 0.81 C_{tG\Phi} \bigr]\,
  (\Lambda/{\rm TeV})^{-2} & 
  at $\sqrt s=1.8$ TeV,\cr\noalign{\vskip 9pt}
 \bigl[  -0.85 (C_{tG}+ C_{qG}) + 1.09 C_{tG\Phi} \bigr]\,
  (\Lambda/{\rm TeV})^{-2} & 
  at $\sqrt s=2$ TeV.\cr}
\end{equation}

\subsection{Current bounds from Run 1}

Current bounds for the coupling strength of the operators can
be derived from the available data on the cross section at Run~1
of the Tevatron.  The production cross section measured by the CDF 
collaboration with an integrated luminosity of 110 pb$^{-1}$ is~\cite{ref14} 
$\sigma=7.6{+1.8\atop-1.5}$ pb for $m_t=175$ GeV, 
which combines dilepton, lepton $+$ jets and all-hadronic channels.  The D0 
collaboration gives~\cite{D0xstt,Klima} $\sigma=5.9\pm1.7$ pb for $m_t=172$ 
GeV.  We use the (unofficial) combined result~\cite{Klima} of 
\begin{equation}
\sigma_{t\bar t}^{\rm exp}=6.7\pm1.3\;\rm pb.
\end{equation}
For the theoretical cross section in the SM, we adopt the most complete  
result currently available~\cite{Bonciani}, which includes soft-gluon 
summation up to the next-to-leading logarithmic order:
\begin{equation}
\textstyle \sigma^{\rm SM}_{t\bar t}= 5.06{+0.13\atop-0.36}\,\rm pb
\end{equation}
for $m_t=175$ GeV at $\sqrt s=1.8$ TeV.
To make a comparison of the measured and standard model cross section, 
we have to take into account the present uncertainty of the top quark 
mass~\cite{CDFmt,D0mt,Klima} which affects both the experimental and 
theoretical cross sections.  Shifting $m_t$ by $\pm5$ GeV changes  
the SM cross section by about $\pm15\%$.  The measured 
cross section also changes with $m_t$ in the same direction but the 
dependence is weaker.  The net effect is about 10\%\ uncertainty in 
the difference of the two cross sections.  

Combining all these uncertainties, the possible new physics contribution 
to the cross section is found to be
\begin{equation}
\Delta\sigma^{\rm new}_{t\bar t} = \sigma^{\rm exp} - \sigma^{\rm SM} 
= 1.6 \pm 1.4 \;\rm pb, 
\end{equation}
which gives a $2\sigma$ bound on the coupling strengths for $\Lambda=1$ TeV
\begin{equation}
-7.2 < C_{tG} + C_{qG} - 1.33 C_{tG\Phi} < 2.0 \;.
\end{equation}
If one of the dimension-six operator gives the dominant 
new physics contribution, we find 
\begin{eqnarray} 
\label{tGb1}
&& -7.2 < C_{tG}, C_{qG} < 2.0 \\
\label{Phib1}
&& -1.5 < C_{tG\Phi} < 5.4 \;.
\end{eqnarray}
We can see that the current bounds from Run~1 are not very strong.
If one defines the new physics scale $\Lambda$ such that the magnitude 
of the coupling strengths are 1, then Run~1 has excluded the existence
of new physics below 400--800 GeV.

\subsection{Expectations for Run~2 and 3}

To determine the size of the couplings that can be probed in the
upgraded Tevatron, we must estimate the number of top events produced.
In our analysis, Run~2 and Run~3 are defined as an integrated luminosity 
of 2 and 30 fb$^{-1}$ at $\sqrt s = 2$ TeV.
We use the SM cross section for $t\bar t$ production of 7.0 pb at 
$\sqrt s=2$ TeV.  
At Run~2 (Run~3), the total number of the produced $t\bar t$ pairs is 
thus about $10^4$ ($2\times 10^5$).  
A detailed analysis of the detection efficiencies and 
signal purity for the three modes, the dilepton ($\ell\ell$), 
lepton plus ${\ge}3$ jets with one $b$-tag ($\ell$3j/$b$), and 
lepton plus ${\ge}4$ jets with two $b$-tags ($\ell$4j/2$b$) can be found
in Ref.~\cite{ref1}. They are shown in Table~\ref{table:xstt}.

In extracting the new physics contribution, various systematic uncertainties 
have to be taken into account besides the experimental systematic error.  
The present uncertainty in the theoretical $t\bar t$ cross section in the 
standard model is at the 5\%\ level~\cite{Bonciani}.  Additional error 
coming from the present error on $m_t$ will be much reduced with the 
expected more precise determination of $m_t$ (2.8 and 0.8 GeV are quoted 
for Run~2 and 3~\cite{ref1}).  We use the total systematic error of 
5\%\ and 1\%\ for illustration.  We also assume  the same detection 
efficiencies as the SM events for the new physics contribution.

Assuming no signal of new physics, the expected $2\sigma$ bound on the
coupling strength of the operators are obtained as shown in
Table~\ref{table:lim}{}; bounds are listed for systematic uncertainties
of both 5\% and 1\%.  A large improvement in the systematic
error does not lead to a corresponding decrease in the bound at Run~2
because much of the error there still comes from the statistical
uncertainty.  At Run~3 the error is dominated by the systematic
uncertainty and the bound would be substantially smaller if the
systematic uncertainty could be reduced to 1\%. For $C_{tG} = C_{qG} =
C_{tG\Phi} = 1$, these bounds correspond to the new physics scale
$\Lambda$ between 1~TeV and 2.5~TeV. Although Run~2 can improve the
bounds from those at Run~1, one needs a better understanding of the
theoretical uncertainty for further significant improvement of the bound
at Run~3.

\section{Disentangling different operators}
\label{sec:4}

In the preceding section we assumed the existence of one operator
at a time and derived the bound for each operator from
the event-number counting experiments at the future runs of the Tevatron. 
But the counting experiments cannot distinguish the effects of different
operators. If the operators coexist, their effects have to 
be disentangled by analyzing additional measurable quantities.
Here we present two methods to distinguish the effects of different 
operators contributing to the $gt\bar t$ coupling; one is the $t\bar t$ 
invariant mass distribution of the cross section, and the other is the 
asymmetry between left- and right-handed top events in top pair production.

\subsection{$t\bar t$ invariant mass distribution}
\label{sec:mtt}

The contribution of  $O_{tG}$ and  $O_{qG}$ is energy dependent 
and thus the behavior of the cross section versus the invariant 
$t\bar t$ mass differs from that predicted by the SM{}.  On the other
hand, the contribution of  $O_{tG\Phi}$ is energy independent 
and gives the same $M(t\bar t)$ distribution as the SM{}.
This provides a method to distinguish the effects of $O_{tG}$ and/or 
$O_{qG}$ from that of $O_{tG\Phi}$.

The mass distribution is given by
\begin{eqnarray}
\frac{d\sigma}{dM_{t\bar t}}&\equiv& 
\frac{d(\sigma^{\rm SM}+\Delta\sigma^{\rm new})}{dM_{t\bar t}} \nonumber\\
&=&\frac{2M_{t\bar t}}{s}
\left (\hat \sigma^{\rm SM}+\Delta\hat \sigma^{\rm new}\right )
\,{dL_{q\bar q}\over d\tau}(\tau=M_{t\bar t}^2/s).
\end{eqnarray}
Figure~1 shows this distribution with and without the new physics
contribution. It can be seen that the presence of $O_{tG\Phi}$ only
alters the magnitude of the cross section, leaving the shape unchanged.  
The effect $O_{tG}$ is energy-dependent and is more prominent for 
larger $M_{t\bar t}$, giving some distortion in the invariant mass 
distribution.  The result for $O_{qG}$ is the same as that for $O_{tG}$ 
and is not shown.   

In order to quantify the shape of the distribution, we define the 
ratio of the high-mass vs. low-mass $t\bar t$ events as follows:
\begin{equation}\label{ratio}
R \equiv \frac{N( M_{t\bar t}>M_0)}{N( M_{t\bar t}<M_0)}
 =\frac{\sigma( M_{t\bar t}>M_0)}{\sigma( M_{t\bar t}<M_0)},
\end{equation}
where $N(M_{t\bar t}>M_0)$ is the number of events
for $M_{t\bar t}>M_0$, {\it etc}.  The value of $M_0$ may be chosen to 
maximize sensitivity to the new physics.  
The ratio $R$ is found to be a function of 
$C_{tG}+C_{qG}$.  It thus provides independent information on new 
physics not contained in the total cross section.  
In particular, $R$ will be the same as the SM prediction if only $O_{tG\Phi}$ 
is present.

Figure~2 shows the ratio $R$ versus the coupling strength 
with and without the presence of  $O_{tG}$.  
We have used $M_0=600$ GeV for illustration.  We note that measuring 
this ratio will not only distinguish $O_{tG}$ (or $O_{qG}$) from $O_{tG\Phi}$
but also determine the sign of the coupling constant $C_{tG}$ (or $C_{qG}$).

\subsection{Top polarization asymmetry} 
\label{sec:asym}

In Sec.~\ref{sec:3}, we have evaluated the new physics contribution to the 
total top pair events summing over the spins of $t$ and $\bar t$.
Measurement of top polarization can give independent information for the 
new physics.  In fact, the axial vector coupling $F_A$ in Eq.~(\ref{gtt}) 
contributes with the opposite sign to left- and right-handed top quark 
cross sections, producing a nonzero polarization asymmetry without 
changing the total cross section. 
 
Following Ref.~\cite{ref15}, we define the asymmetry as
\begin{equation}\label{Asy}
{\cal A} \equiv \frac{ N_{R} -N_{L} }{ N_{R} +N_{L} } 
= \frac{ \sigma_{R} -\sigma_{L} }{ \sigma_{R} +\sigma_{L} },
\end{equation}
where $N_R$ and $N_L$ are the number of right- and left-handed top
quarks, respectively. We do not require spin information on the
$\bar t$. The polarization may be measured%
\footnote{Anomalous $tbW$ couplings may change the decay angular 
distribution.  However, the three operators considered in this and the 
previous sections do not produce such couplings.} 
through the angular distributions~\cite{ref16} of the leptonic events 
$t \to W^+ b \to \ell^+\nu_\ell b$ ($\ell=e, \mu$).  
In the SM this asymmetry is too small to be observed at the Tevatron, 
because QCD is invariant under parity and charge conjugation, 
and so the asymmetry arises only from the weak corrections~\cite{ref15}.
Hence the asymmetry is a good observable for probing new physics.

The cross sections in (\ref{Asy}) can be written as 
$\sigma_{R} = \sigma_{RL} +\sigma_{RR}$ and 
$\sigma_{L} = \sigma_{LR} +\sigma_{LL}$, with 
$\sigma_{\lambda_1\lambda_2}\equiv 
\sigma( p\bar{p} \to t_{\lambda_1} \bar{t}_{\lambda_2}+X)$.  
Here $\lambda_1$ and $\lambda_2$ indicate the helicity states for $t$ 
and $\bar t$, respectively.  These are obtained by convoluting 
the parton-level cross section $\hat{\sigma}(q\bar{q} \to t\bar{t})$
with parton distribution functions as in Eq.~(\ref{eq22}). 
The parton-level cross section is expressed as 
\begin{equation}
\hat \sigma_{\lambda_1\lambda_2}=\hat \sigma_{\lambda_1\lambda_2}^{\rm SM}
                       +\Delta \hat\sigma_{\lambda_1\lambda_2}^{\rm new},
\end{equation}
where $\hat \sigma_{\lambda_1\lambda_2}^{\rm SM}$ and
$\Delta \hat\sigma_{\lambda_1\lambda_2}^{\rm new}$ are the SM and new 
physics contribution, respectively.  At tree level, these are given by 
\begin{mathletters}
\begin{eqnarray}
\hat{\sigma}^{\rm SM}_{LL} & = & \hat{\sigma}^{\rm SM}_{RR}
 = \frac{ 4 \pi \alpha_s^2 \beta}{27\hat s}\,{2m_t^2\over \hat s}, \\
\hat{\sigma}^{\rm SM}_{LR} & = & \hat{\sigma}^{\rm SM}_{RL}
 = \frac{ 4 \pi \alpha_s^2 \beta}{27\hat s},
\end{eqnarray}
\end{mathletters}
and
\begin{mathletters}
\begin{eqnarray}
\Delta\hat \sigma_{LL}^{\rm new}&=&\hat \sigma_{RR}^{\rm new}
                         =\frac{1}{g_s}
\left ( 2F_V +\frac{\beta^2 \hat{s}}{2m_t^2} F_M \right )
                          \hat{\sigma}^{\rm SM}_{LL},\\
\Delta\hat \sigma_{LR}^{\rm new}&=&\frac{2}{g_s}
                         \left ( F_V -\beta F_A \right )
                         \hat{\sigma}^{\rm SM}_{LR},\\
\Delta\hat \sigma_{RL}^{\rm new}&=&\frac{2}{g_s}
                         \left ( F_V+\beta F_A \right )
                         \hat{\sigma}^{\rm SM}_{RL}.
\end{eqnarray}
\end{mathletters}
Here we see that $\Delta\hat \sigma_{LR}^{\rm new}$ differs 
from $\Delta\hat \sigma_{RL}^{\rm new}$ due to the existence of $F_A$,
which will cause the asymmetry ${\cal A}$. 

To order $1/\Lambda^2$, we can neglect the new physics 
contribution to the denominator in Eq.~(\ref{Asy}). 
Again we use the CTEQ3L parton distribution functions~\cite{ref13}  
with $\mu=\sqrt {\hat s}$. For  $m_t=175$ GeV, we obtain the asymmetry 
\begin{equation}
{\cal A}=-0.098 (C_{tG}-C_{qG})\,(\Lambda/{\rm TeV})^{-2}.
\end{equation}
A cut on the $t\bar t$ invariant mass $M_{t\bar t}>M_0$ enhances the 
asymmetry but at the same time decreases the number of $t \bar t$ events.

The expected number of reconstructed $t \bar t$ events with top decay 
in the channel $t \to W^+ b \to \ell^+\nu_\ell b$ ($\ell=e, \mu$) may 
be found in Table~\ref{table:xstt}, which implies 
the 1$\sigma$ sensitivity for the asymmetry of $\sim 3 \%$ and 
$\sim 1 \%$ in magnitude at Run~2 and Run~3, respectively.  
Under the present constraints Eq.~(\ref{tGb1}), 
the operators $C_{tG}$ or $C_{qG}$ can produce an asymmetry ${\cal A}$
as large as 30\%\ assuming no cancellation between these two operators.  
Such a large asymmetry should be clearly observable in future runs.  
If no asymmetry is observed, one can put bounds on the operators
\begin{equation}\label{asyb}
\frac{\vert C_{tG}-C_{qG}\vert }{(\Lambda/{\rm TeV})^2} \le 1.3 \;(0.4).
\end{equation}
for Run~2 (Run~3), which corresponds to ${\cal A} \leq 4/\sqrt N$ ($N$
is the total number of events).  Although this bound is weaker than that
obtained in Sec.~\ref{sec:3}, it can be regarded as independent
information because it does not depend on the coupling strength of
$O_{tG\Phi}$.

Note that both of the observables in Sec.~\ref{sec:mtt}\ and
Sec.~\ref{sec:asym} distinguish the effects of $O_{tG}$ or $O_{qG}$ from
that of $O_{tG\Phi}$.  Furthermore, the two methods are complementary in
that the ratio $R$ in Eq.~(\ref{ratio}) depends on $C_{tG}+C_{qG}$ while
the asymmetry ${\cal A}$ in Eq.~(\ref{Asy}) is sensitive to
$C_{tG}-C_{qG}$.  Therefore, separation of the effects of the three
couplings is possible using the measurements discussed in this and the
previous sections.

\section{Single top quark production}
\label{sec:5}

Now we examine the effects of the set of operators 
(\ref{OqW})--(\ref{OtWPhi}) which contribute to single top production
At the Tevatron, this reaction occurs mainly through 
the $s$-channel $W^*$ process $q'\bar q\rightarrow t\bar b$ and
the $W$-gluon fusion process, which were studied extensively in the 
SM~\cite{ref17} and some of its extensions~\cite{ref18}.    
The $s$-channel $W^*$ process, despite its relatively low cross section,  
is quite powerful for probing new physics because (i) the systematic 
error in the theoretical calculation of its cross section is small
(its initial state effects can be measured in the similar Drell-Yan
process $q'\bar q\rightarrow \ell\nu$), (ii) it can be isolated from other
single top production processes at the upgraded Tevatron by requiring 
that both jets in the final state be tagged as $b$ jets~\cite{ref19}. 
The balance between statistics and systematics for the  $W^*$ process
gives the result that its cross section can be measured to about the
same precision as that for full single top cross section.
So we choose the $s$-channel $W^*$ process to probe the operators.%
\footnote{The anomalous $Wtb$ couplings may also be probed
at the Tevatron in the top quark decay $t\rightarrow W b$ 
with spin analysis~\cite{Nelson}.}

Unlike the case of top pair production in which the efficiency for 
top selection can be predicted by extrapolating from Run~1 experience,
we will calculate the number of single top signal and backgrounds
by Monte Carlo simulation.
We will also take into account of the fact that the top quark is polarized 
in its production.

\subsection{Signal and backgrounds}

For both $t$ and $b$ being on-shell, the new physics contribution
to the $Wt\bar b$ vertex can be written as 
\begin{eqnarray}
\Gamma^{\mu}_{Wt\bar b}&=&-\frac{g_2}{\sqrt 2}\left [
  \gamma^{\mu} \bigl( \kappa_{1L} P_L + \kappa_{1R} P_R\bigr)
 + p_t^{\mu} \bigl( \kappa_{2L} P_L + \kappa_{2R} P_R\bigr)
 + p_{\bar b}^{\mu} \bigl( \kappa_{3L} P_L + \kappa_{3R} P_R\bigr) \right ],
\end{eqnarray}
where $P_{L,R}\equiv(1\mp \gamma_5)/2$ and the form factors from new physics
are given by
\begin{mathletters}
\begin{eqnarray}
\kappa_{1L}&=&\frac{v^2}{\Lambda^2}\left [ 
 C_{tW\Phi}\frac{\sqrt 2 m_t}{g_2 v}+C^{(3)}_{\Phi q}
   -C_{qW}\frac{k^2}{g_2v^2} \right ],\\
\label{rhand}
\kappa_{1R}&=&\frac{v^2}{\Lambda^2}\left [ 
 C_{bW\Phi}\frac{\sqrt 2 m_t}{g_2 v}+\frac{C_{t3}}{2} \right ],\\
\kappa_{2L}&=&\frac{v}{\Lambda^2}\left [ 
 -C_{tW\Phi}\frac{\sqrt 2 }{g_2}-\frac{C_{Dt}}{\sqrt 2}
   +C_{qW}\frac{m_t}{g_2 v} \right ],\\
\kappa_{2R}&=&-\frac{v}{\Lambda^2}
 C_{bW\Phi}\frac{\sqrt 2}{g_2},\\
\kappa_{3L}&=&\frac{v}{\Lambda^2}\left [ 
 C_{tW\Phi}\frac{\sqrt 2 }{g_2}
   +C_{qW}\frac{m_t}{g_2 v} \right ],\\
\kappa_{3R}&=&\frac{v}{\Lambda^2}\left [
\frac{C_{Db}}{\sqrt 2} + C_{bW\Phi}\frac{\sqrt 2}{g_2} \right ],
\end{eqnarray}
\end{mathletters}
where $k=p_t+p_{\bar b}$. 

The interference of the SM matrix element (${\cal M}_{\rm SM}$) 
with the new physics contribution (${\cal M}_{\rm new}$) 
for the $W^*$ process $u+\bar d\to t+\bar b$ is given by
(neglecting KM mixing)
\begin{equation}
\label{eq:xsst}
\sum_{\rm spins}
\mathop{\rm Re} \bigl[ {\cal M}_{\rm SM}^{\dagger} {\cal M}_{\rm new} \bigr] 
= {g_2^4\over 2(\hat s-m_W^2)^2}\,\bigl[ 2\hat u(\hat u-m_t^2) \kappa_{1L} 
-  m_t \hat t\hat u (\kappa_{2L}-\kappa_{3L}) \bigr]
\end{equation}
for spin-summed matrix elements and 
\begin{equation}
\mathop{\rm Re} \bigl[ {\cal M}_{\rm SM}^{\dagger} {\cal M}_{\rm new} \bigr] 
= \cases{\displaystyle
{g_2^4\hat t\hat u \over 2(\hat s-m_W^2)^2}\,
\biggl[ {2m_t^2\over \hat s-m_t^2} \,\kappa_{1L}
-m_t (\kappa_{2L}-\kappa_{3L}) \biggr] & for $h_t=+$\cr
\noalign{\vskip 9pt}\displaystyle
{g_2^4\over (\hat s-m_W^2)^2}\,{\hat s\hat u^2\over \hat s-m_t^2}\,\kappa_{1L}
& for $h_t=-$}
\end{equation}
for matrix elements with top quark helicity $h_t=\pm$.  Here $\hat s$ is 
the parton c.m.\ energy squared and
\begin{mathletters}
\begin{eqnarray}
&&\hat t = (p_u-p_t)^2 = -{1\over2}(\hat s-m_t^2)(1-\cos\theta^*)\\
&&\hat u = (p_d-p_t)^2 = -{1\over2}(\hat s-m_t^2)(1+\cos\theta^*)
\end{eqnarray}
\end{mathletters}
where $\theta^*$ is the c.m.\ scattering angle.  We have neglected the 
bottom quark mass.  With this approximation, only one helicity of the 
quarks other than the top participate in the process and 
the contribution of $\kappa_{iR}$ ($i=1$, 2, 3), and hence that of 
the operators $O_{bW\Phi}$, $O_{t3}$, and $O_{Db}$, drops out.

For the $W^*$ process, we look for events with
 $t\rightarrow W^+b \rightarrow \ell^+\nu b$ ($\ell=e, \mu$)
and thus the signature is an energetic
charged lepton, missing $E_{T}$, and  double $b$-quark jets.
We assumed silicon vertex tagging of the $b$-quark jet 
with $50 \%$ efficiency and the probability of 0.4\% for a light quark 
jet to be misidentified as a $b$-jet. 
The potential SM backgrounds are:
(B1) the same $W^*$ process in the SM,
(B2)  the quark-gluon fusion process $qg\rightarrow q't\bar b$ where
$q^\prime$ is misidentified as a $b$-jet, 
(B3)  processes involving a b-quark in the initial state,
     $qb \rightarrow tq'$ and $gb\rightarrow tW$,
(B4) $t\bar t\rightarrow W^-W^+b\bar b$.
(B5)   $Wb\bar b$, and  
(B6)   $Wjj$.

Let us discuss the backgrounds in turn. Background process (B2) contains an
extra light quark jet and can only mimic our signal 
if the quark misses detection by going into the beam 
pipe. In our calculation of the $W$-gluon fusion process as a background, we
impose $\eta(q') >3$ and $p_T(q')< 10$ GeV for the  light-quark jet.
The $qb \rightarrow tq'$ background%
\footnote{There is a large uncertainty in the relative
strengths of the $qb$ and $qg$ contributions to the background
since the $b$ quark parton distribution is not a measured quantity, but
is rather an entirely theoretical construction. 
Since both the $qb$ and $qg$ contributions after all
cuts are a very small fraction of the total background, this uncertainty
will not affect our results.} is greatly 
reduced by requiring double $b$-tagging. The process $gb\rightarrow tW$
can only imitate our signal if the $W$ decays into two jets, where 
one jet is missed by the detector and the other is misidentified as a
$b$ quark, which should be negligible for the misidentification rate
assumed here. 
Background process (B4) can mimic our signal if both $W$'s decay leptonically
and one charged lepton is not detected, which we assumed to occur if
$\vert \eta(\ell) \vert >3$ and $p_T(\ell)<10$ GeV.
Since we required two $b$-jets to be present in the final state,
the potentially large background process (B5) from $Wjj$ is reduced 
to an insignificant level. In order to reduce the backgrounds 
$Wb\bar b$ and $Wjj$, we required the reconstructed top quark mass 
$M(bW)$ to lie within the mass range 
$\left \vert M(bW)-m_t\right \vert <30$ GeV.

 To simulate the detector acceptance,  we made a series of
cuts on the transverse momentum ($p_{T}$), the pseudo-rapidity
($\eta$), and the separation in the azimuthal angle-pseudo rapidity plane 
($\Delta R= \sqrt{(\Delta \phi)^2 + (\Delta \eta)^2} )$ between a jet 
and a lepton or between two jets. The cuts are chosen to be
\begin{mathletters}
\begin{eqnarray}
&& p_T^\ell, \; p_T^b,\;
p_T^{\rm miss}\ge 20 \rm{~GeV} ,\\
&& \vert\eta_{b}\vert,\; \vert\eta_{\ell}\vert \le 2.5 ,\\
&& \Delta R_{jj},\; \Delta R_{j\ell} \ge 0.5 .
\end{eqnarray}
\end{mathletters}
To make the analyses more realistic, we simulate the detector effects
by  assuming a Gaussian smearing from the energy of the final state
particles, given by:
\begin{equation}
\Delta E / E  = \cases{
30 \% / \sqrt{E} \oplus 1 \% & for leptons,\cr
80 \% / \sqrt{E} \oplus 5 \% & for hadrons,\cr}
\end{equation}
where $\oplus$ indicates that the energy dependent and independent
terms are added in quadrature and $E$ is in GeV.

We have explicitly calculated backgrounds (B1) and (B2), and for the
others  used the analysis of Ref.~\cite{ref20}. 
After all cuts, the number of background events 
at Run~2 (Run~3) are found to be 38 (570), 2 (26), 1 (14),
4 (54), 60 (900) and 1 (21), respectively, for the six background
processes described above. Here we see that after the cuts the backgrounds
from  processes (B2), (B3), (B4) and (B6) are negligibly small.
The number of total background events is 106 (1590) for Run~2 (Run~3).  
After the cuts, the number of signal events from new physics is found to be 
\begin{equation}
S=\cases{
   \bigl( 9 C_{tW\Phi} + 5 C_{\Phi q}^{(3)} - 11 C_{qW} +0.3 C_{Dt}\bigr)
   \bigl(\Lambda/(\rm TeV)\bigr)^{-2}, & at Run~2\cr
\noalign{\vskip 6pt}
   \bigl( 135 C_{tW\Phi} +69 C_{\Phi q}^{(3)} - 165 C_{qW} +4.5 C_{Dt}\bigr)
   \bigl(\Lambda/(\rm TeV)\bigr)^{-2}, & at Run~3.\cr}
\end{equation}
The effects of $O_{Dt}$ are much smaller than those of the other three
operators since it does not contribute to the form factor $\kappa_{1L}$,
whose contribution is found to be much larger than those of the other
form factors.

\subsection{Improvement of the bounds with Run~2 and 3}

From the results of the preceeding subsection, 
we obtain the bounds on the coupling strength 
of the operators from Run~2 (Run~3) if the new physics events are 
not observed at the $2\sigma$ level
\begin{mathletters}\label{qwnetc}
\begin{eqnarray}
\frac{\vert C_{qW}\vert }{(\Lambda/{\rm TeV})^2}   &\le& 2.1~(0.50), \\
\frac{\vert C_{\Phi q}^{(3)}\vert }{(\Lambda/{\rm TeV})^2}&\le& 4.6~(1.2), \\
\frac{\vert C_{tW\Phi}\vert }{(\Lambda/{\rm TeV})^2}&\le& 2.6~(0.61), \\
\frac{\vert C_{Dt}\vert }{(\Lambda/{\rm TeV})^2}&\le& 77~(18),
\end{eqnarray}
\end{mathletters} 
where we again assumed the simple situation that cancellation among different 
operators does not take place.
 
As mentioned in Sec.~\ref{sec:2}, the operators 
$O_{qW}$, $O_{\Phi q}^{(3)}$, $O_{Db}$ and
$O_{bW\Phi}$ also affect the $Zb\bar b$ coupling
and will be subject to an $R_b$ constraint.
The SM values of $R_b$ and the latest experimental data are~\cite{ref21} 
\begin{equation}\label{data1}
R_b^{\rm SM} = 0.2158, \qquad R_b^{\rm exp} = 0.2170(9).
\end{equation}
With the new physics contribution described in Eq.~(\ref{zbb}), 
$R_b$ is given by
\begin{equation}\label{Rb}
R_b=R_b^{SM}\left[ 1+ \frac{4s_Wc_W}{e}\frac{vm_Z}{\Lambda^2}
                   \left ( C_{qW}\frac{c_Wm_Z}{2v}-C^{(3)}_{\Phi q} \right )
                   \frac{v_b+a_b}{v_b^2+a_b^2}(1-R_b^{SM})\right ],
\end{equation}
where we have neglected the bottom quark mass and thus 
the contributions of  $O_{bW\Phi}$ and $O_{Db}$, which 
are proportional to $m_b/m_Z$, drop out.
At the $2\sigma$ level, we obtain the bounds for the coupling strength
by assuming that no cancellation between $O_{qW}$ and $O^{(3)}_{\Phi q}$  
takes place, 
\begin{mathletters}\label{qwetc}
\begin{eqnarray}
-0.8 &<&\frac{C_{qW}}{(\Lambda/{\rm TeV})^2}<0.2,\\
-0.03 &<&\frac{C^{(3)}_{\Phi q}}{(\Lambda/{\rm TeV})^2}<0.13.
\end{eqnarray}
\end{mathletters}

Since the dimension-six operators give a bad high energy behavior, 
there is an energy scale above which they cease to be a valid description 
of new physics.  Any process below the new physics scale $\Lambda$ should 
not violate the unitarity limit.   For a given $\Lambda$, this requirement 
can be translated to an upper limit on the coupling strengths $C$.
These unitarity limits have been worked out by Gounaris 
{\it et al.}~\cite{Gounaris2} 
in detail.  
For $O_{Dt}$ and $O_{tW\Phi}$, the strongest limits are obtained from 
two-body $t\bar t$ scattering process.  For $\Lambda=1$ TeV, they are given 
by
\begin{mathletters}\label{twphietc}
\begin{eqnarray}
 \vert C_{tW\Phi} \vert &<&13.5,\\
 \vert C_{Dt} \vert &<&    10.4.
\end{eqnarray}
\end{mathletters}
The limit on $C_{Dt}$ is independent of $\Lambda$.  That on $C_{tW\Phi}$ 
becomes somewhat weaker for larger $\Lambda$.  

Comparing Eqs.~(\ref{qwnetc}) with Eqs.~(\ref{qwetc}) and 
(\ref{twphietc}), we find:
\begin{enumerate}
\item For the operators $O_{qW}$ and $O^{(3)}_{\Phi q}$, 
                 future runs at the Tevatron cannot improve the current
                 bounds obtained from $R_b$, which place much stronger
                 constraints on these operators. Hence their effects
                 will not be observable at the Tevatron, even for a 
                 luminosity of 100 fb$^{-1}$.
\item  The operator $O_{tW\Phi}$, currently subject only to
                  weak bounds from unitarity, can be meaningfully 
                  probed at Run~2 and Run~3. 
\item  The operator $O_{Dt}$ cannot be probed at Run~2 and
                 Run~3 much beyond the current bound from unitarity.
                 For a higher integrated luminosity of 100 fb$^{-1}$,
                 we found that the bound is  $C_{Dt}<9.8$ which is
                 only slightly stronger than its current bound.
\end{enumerate}
So we conclude that among the operators contributing to the $Wt\bar b$
coupling, only one operator ($O_{tW\Phi}$) can be meaningfully 
probed at the future runs of the Tevatron.

\section{Conclusions}
\label{sec:last}

We have studied in the effective Lagrangian approach the ability of
future runs at the Tevatron in probing anomalous couplings of the top quark.
We have listed and analyzed the possible dimension-six CP-conserving
operators involving the anomalous $gt\bar t$ or $Wt\bar b$ couplings
which could be generated by new physics at a higher scale.

For the operators which give rise to an anomalous $gt\bar t$ coupling,
we evaluated their effects in top pair production and derive the 
bounds both from Run~1 and those expected from future runs.
We found that the current constraints from Run~1 are not very strong 
and that future runs can either discover the effects of these
operators or significantly improve the current bounds.
We also proposed two methods to disentangle the effects between different 
operators contributing to the $gt\bar t$ coupling:
one is studying the energy distribution of the 
cross section and the other is measuring the asymmetry between 
left- and right-handed top events in top pair production.

For the operators which give rise to an anomalous $Wt\bar b$ coupling, we
calculated their contribution to single top production at the upgraded
Tevatron and derived the bounds which could be obtained from Run~2 and
Run~3.  We found that future runs of the Tevatron cannot effectively
probe those operators which are currently subject to a tight constraint
from $R_b$. For the operators which are not
subject to the $R_b$ constraint and are only constrained by unitarity, 
future runs of the Tevatron can
either discover the effects of or set stronger bounds on
$O_{tW\Phi}$, while the best bound on $O_{Dt}$ is still obtained
from the unitarity constraint for any new physics scale.  

In Table~\ref{table1} we summarize the strongest upper bounds on the
operators under consideration that currently exist or can be obtained in
the future at the Tevatron, for $\Lambda=$ 1 TeV. The current bounds are
from our results using either $R_b$ or the Run~1 data at the $2\sigma$
level, except for $C_{tW\Phi}$ and $C_{Dt}$ where the best current bounds
come from the requirement of unitarity.
We note that the current bounds on $C_{qW}$ and
$C^{(3)}_{\Phi q}$ obtained from $R_b$ are better than those that can be
obtained at Run~3 at the Tevatron.

\section*{Acknowledgments}

J.~M.~Y. acknowledges JSPS for the Postdoctoral Fellowship.  
The work of K.~H. and J.~M.~Y. is supported in part by the 
Grant-in-Aid for Scientific Research (No.~10640243) and Grant-in-Aid 
for JSPS Fellows (No.~97317) from the Japan Ministry of Education, 
Science, Sports, and Culture.  This work was supported in part 
by the U.S. Department of Energy, Division
of High Energy Physics, under Grant No. DE-FG02-94ER40817.


\begin{table}
\caption{Expected numbers of events and statistical errors for 
top pair production at Run 2 and 3.  The combined statistical error 
is derived from the $\ell\ell$ and $\ell$3j/$b$ channels.}
\label{table:xstt}
\vspace{7mm}
\begin{tabular}{lcccccc}
  & & & \rlap{\kern8mm Run 2} & & \rlap{\kern8mm Run 3} &\\
Mode & Efficiency  & Purity & \#events  & Statistical
 & \#events  & Statistical\\
&(\%)& S:B &(with bkgd) & error (\%) & (with bkgd) & error (\%)\\\hline
$\ell\ell$    & 1.2  & 5:1   & 200  & 8.5  & 3000  & 2.2\\
$\ell$3j/$b$  & 8.6  & 3:1   & 1600 & 3.3  & 24000 & 0.9\\
$\ell$4j/2$b$ & 3.8  & 12:1  & 570  & 4.5  & 8600  & 1.2\\
total         &      &       &      & 3.1  &       & 0.8\\
\end{tabular}
\end{table}

\begin{table}
\caption{Expected 2$\sigma$ bounds at Run 2 and 3 
for the operators contributing top pair production.  If the three operators 
coexist, the limits for $|C_{tG}|$ {\it etc.\/} should be reinterpreted 
to those for $|C_{tG}+C_{qG}- 1.28 C_{tG\Phi}|$.}
\label{table:lim}
\vspace{7mm}
\begin{tabular}{lcccc}
                  & \rlap{\kern12mm Run 2} &&\rlap{\kern12mm Run3}&\\
Systematic error  & 5\%  & 1\%  & 5\%  & 1\% \\\hline
$|C_{tG}|$, $|C_{qG}|$ & 0.97 & 0.53 & 0.82 & 0.21 \\
$|C_{tG\Phi}|$      & 0.75 & 0.41 & 0.64 & 0.17 \\
\end{tabular}
\end{table}

\begin{table}
\caption{
Current and future strongest upper bounds on the magnitude 
of coupling strength 
of some operators at $2\sigma$ level for $\Lambda=$ 1 TeV.      
The current bounds  are from unitarity requirement for 
$C_{tW\Phi}$ and $C_{Dt}$, from $R_b$ for $C_{qW}$ and $ C_{\Phi q}^{(3)}$,
and from Run~1 data for $C_{tG}$, $ C_{qG}$ and $C_{tG\Phi}$.}
\label{table1}
\vspace{10mm}
\begin{tabular}{lrrr}
& Current bounds & Bounds from Run~2 &  Bounds from Run~3 \\\hline
$\vert C_{tG}\vert, ~\vert C_{qG}\vert $& 7.2& 1.0 & 0.8   \\ 
$\vert C_{tG\Phi}\vert$                   & 5.4& 0.8 & 0.6 \\ 
$\vert C_{qW}\vert$                       & 0.2& 2.1 & 0.5 \\ 
$\vert C_{\Phi q}^{(3)}\vert$             &0.0\rlap{3}& 4.6 & 1.2 \\ 
$\vert C_{tW\Phi}\vert$                   &13.5& 2.6 &0.6  \\ 
$\vert C_{Dt}\vert$                       &10.4& 77\phantom{.0} 
 &18\phantom{.0}   \\ 
\end{tabular}
\end{table}

\begin{figure}
\vfil
\begin{center}
\psfig{figure=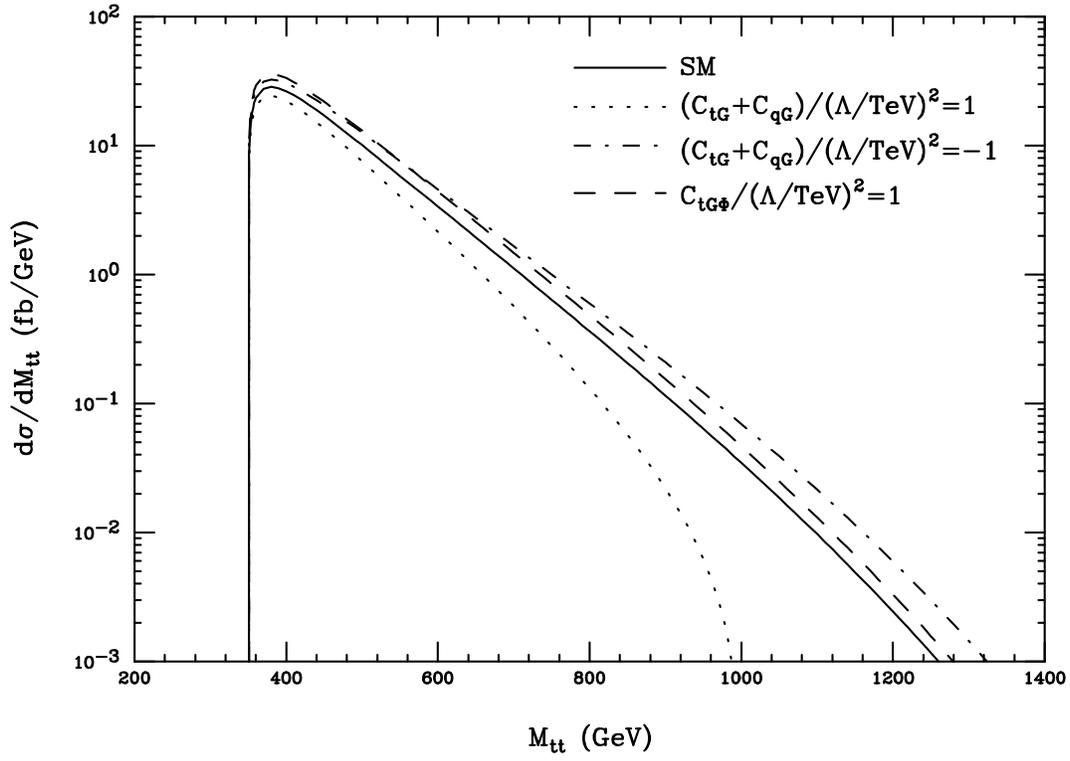,width=400pt,angle=90}
\end{center}
\hspace*{10mm}
\caption{Top pair invariant mass distribution in the SM (solid), 
with contributions from $O_{tG\Phi}$ (dashed) and with $O_{tG}$ 
and/or $O_{qG}$ (dotted and dash-dotted). 
The two curves for $O_{tG}/O_{qG}$ correspond
to different coupling strengths.}
\label{fig:mtt}
\vfil
\end{figure}
\eject\hrule height 0pt
\begin{figure}
\vfil
\begin{center}
\psfig{figure=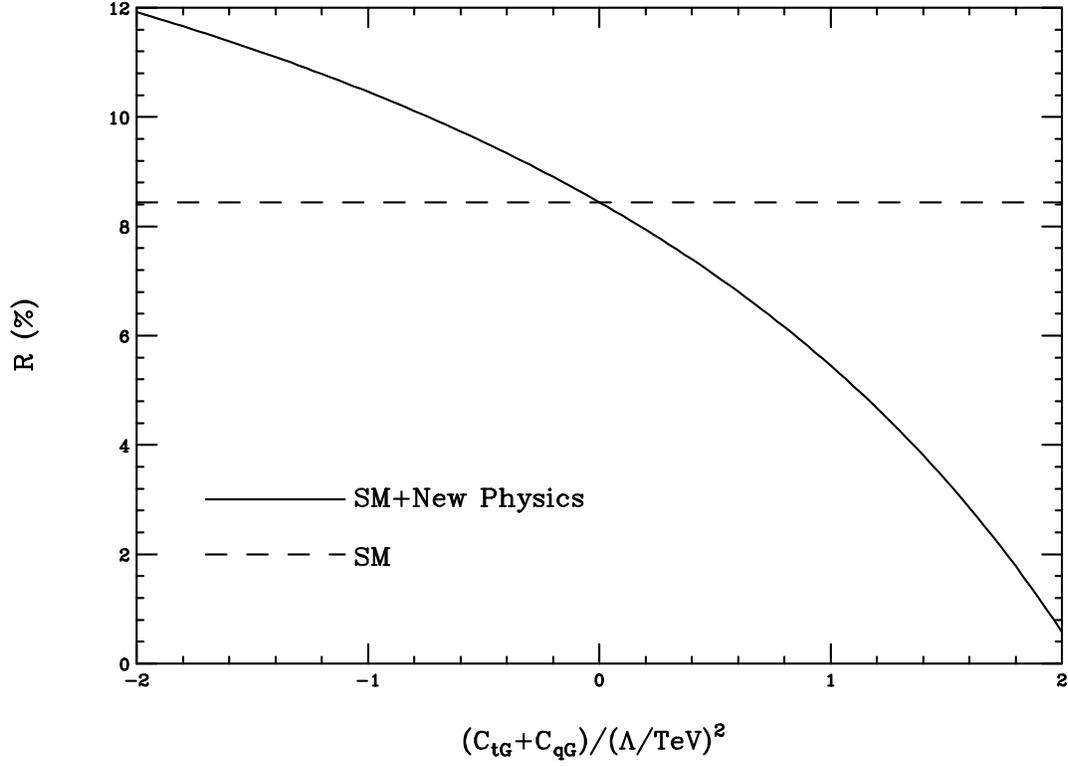,width=400pt,angle=90}
\end{center}
\hspace*{10mm}
\caption{The ratio $R$ with $M_0=600$ GeV as a function of 
$(C_{tG}+C_{qG})/\Lambda^2$.  The SM value is indicated by the dashed line.}
\label{fig:R}
\vfil
\end{figure}
\end{document}